\documentstyle[astrobib,epsfig]{mn}

\begin{document}
\def\bi#1{\hbox{\boldmath{$#1$}}}

\newcommand{\beq}{\begin{equation}}
\newcommand{\eeq}{\end{equation}}
\newcommand{\beqa}{\begin{eqnarray}}
\newcommand{\eeqa}{\end{eqnarray}}

\newcommand{\lexp}{\mathop{\langle}}
\newcommand{\rexp}{\mathop{\rangle}}
\newcommand{\rexpc}{\mathop{\rangle_c}}

\def\bi#1{\hbox{\boldmath{$#1$}}}

\title{Cosmological constraints from the masses and abundances of $L_*$
galaxies}

\author[U. Seljak]{
 U.~Seljak\thanks{E-mail: uros@feynman.princeton.edu} \\
 Department of Physics, Princeton University, Princeton, NJ 08544,
 USA}

\pubyear{2001}

\maketitle

\begin{abstract}

We place limits on the mean density of the
universe $\Omega_m$ 
and the effective slope of the linear power spectrum around a megaparsec scale
$n_{\rm eff}$
by comparing the universal mass function to the observed luminosity function.
Numerical simulations suggest that the dark matter halo mass function 
at small scales depends only on $\Omega_m(n_{\rm eff}+3)$
independent of the overall power spectrum normalization.
Matching the halo abundance to the observed luminosity function 
requires knowing the relation between the virial mass and
luminosity (separately for early and late type galaxies) and the 
fraction of galaxies that reside in larger halos such as groups and 
clusters, all  
of which can be extracted from the galaxy-galaxy lensing. 
We apply the recently derived values from SDSS and find
$\Omega_m(n_{\rm eff}+3)= (0.15 \pm 0.05)/(1-f_{\rm dh})$, where 
$f_{\rm dh}$
accounts for the possibility that some
fraction of halos may be dark or without bright central galaxy.
A model with $\Omega_m=0.25$ and primordial $n=0.8$ 
or with $\Omega_m=0.2$ and $n=1$
agrees well with these constraints even in the absence of dark halos, 
although with the current data somewhat 
higher values for $\Omega_m$ and $n$ are also acceptable.
\end{abstract}



\section{Introduction}

Halo mass function has been long recognized as a powerful probe of 
cosmology since the seminal work by \citeN{1974ApJ...187..425P}. 
Most of the applications so far have focused on 
clusters, which are easy to detect  
in X-rays and for which the observed X-ray 
temperatures correlate well with the cluster mass. While there is  
currently some uncertainty in the normalization of this relation, upcoming
X-ray and weak lensing observations should provide empirical means to 
calibrate it.
Since the 
clusters are the most massive halos formed in the universe they 
lie on the exponential tail of the mass function, whose amplitude
depends mainly on the overall normalization of the power spectrum and  
the density parameter $\Omega_m$.
On the other hand, for the halo mass below the nonlinear mass 
the mass function depends only 
on the density of the universe and on the effective slope of the 
linear power spectrum at that scale and is independent of the power 
spectrum normalization 
(\citeNP{1974ApJ...187..425P}, 
\citeNP{1999MNRAS.308..119S}, \shortciteNP{2001MNRAS.321..372J}). 

To constrain these cosmological parameters 
with mass function one must be able to determine
masses and abundances of halos below the cluster scale.
Groups with masses between $10^{13}-10^{14}M_{\sun}$
are difficult to observe directly, since they contain only a 
handful of galaxies, and their abundance is quite 
uncertain in this range. 
Below $10^{11}M_{\sun}$ many of the halos 
may not host a bright galaxy because of effects that prevent 
either gas cooling or star formation, such as 
UV background radiation, feedback or  
insufficient surface density for star formation.
Moreover, a significant fraction of galaxies
corresponding to these halo masses may be satellites inside a larger
halo and one must correct for that, since the halo mass function only 
counts isolated halos. Estimating this fraction  
is difficult for small halos. 

In the mass range 
$10^{11}-10^{12}M_{\sun}$, typical for $L_*$ galaxies, the luminosity 
function is well determined and theoretical models 
suggest that each of these halos should host a bright galaxy 
at the center (e.g. \shortciteNP{1999MNRAS.303..188K}, \shortciteNP{2000MNRAS.311..793B}). The main challenge in this range is to determine
the relation between a
galaxy luminosity and a halo mass and the fraction of these galaxies
that belong to larger halos such as groups and clusters. 
Even though we have good dynamical probes of mass within the 
optical region of a galaxy, such as Tully-Fisher relation for 
late types and 
Faber-Jackson relation or strong lensing 
for early types, the relation 
between optical masses and virial masses  
is more uncertain, since 
the virial mass depends on the adopted density profile and
a typical virial radius is a factor of 10 larger than optical radius.
Virial masses thus cannot be observationally constrained from the optical or
HI studies alone, which was used in previous work on the mass function 
determination on galactic scales (\shortciteNP{2000ApJ...528..145G}, 
\citeNP{2001astro.ph..8160K}).

An alternative approach adopted here is to use virial 
masses derived directly from the galaxy-galaxy (g-g) lensing. In this method 
one uses tangential distortions of background galaxies by the foreground 
galaxy to place limits on the mass distributions around galaxies. 
A recent SDSS  analysis presents galaxy-galaxy lensing  results
using a much larger sample than previously available, 
allowing a detailed study 
of the relation between mass and light for several luminosity bands and 
morphological types
\shortcite{2001astro.ph..8013M}. 
The latter is particularly important since 
the relation between mass and luminosity differs significantly 
between early and late type galaxies and one must extract the
relations separately before 
combining them together. 
The data are most
sensitive to 100-200$h^{-1}$kpc transverse separations, which is a typical
virial radius of a $10^{11}-10^{12}h^{-1}M_{\sun}$ halo.
Second complication is the fraction of 
galaxies that are not in isolated halos. This can also be 
extracted from g-g lensing: if there is a significant lensing signal around
the galaxies at separations above
200$h^{-1}$kpc then it signals a presence 
of groups and clusters around them \cite{2002astro.ph..1448G}.
This allows
one to determine the fraction of galaxies of a given luminosity in these 
larger halos.
Together thus g-g lensing provides all the necessary information 
needed for a quantitative study of halo mass function on galactic scales.

It is worth comparing this approach to the traditional mass to light ratio 
($M/L$) method to obtain the density parameter $\Omega_m$ (e.g. 
\shortciteNP{1995ApJ...447L..81B},
\shortciteNP{1997ApJ...478..462C}). 
This often assumes that $M/L$ 
extracted from some type of objects, such as groups or clusters, can 
be applied to 
the global $M/L$. Since we can measure the total luminosity density in 
the universe we can obtain 
the total matter density by multiplying the two. 
However, $M/L$ 
depends on both the halo mass and scale on which one measures it. While 
light is concentrated to the inner parts of the clusters mass 
continues to increase, so $M/L$ for any given object increases with radius. 
Even adopting the virial
masses, defined so that only baryons within that 
radius can condense to make stars, there is no 
reason why $M_{\rm vir}/L$ should not depend on $L$. 
Theoretical models predict 
$M_{\rm vir}/L$ to be at the minimum at galactic masses, where cooling and star 
formation are most efficient 
(e.g. \shortciteNP{1999MNRAS.303..188K}, \shortciteNP{2000MNRAS.311..793B}). 
It is possible that there are many small halos that have 
no light at all (e.g. \citeNP{1997MNRAS.292L...5J}), which is why we cannot see them, but which contribute 
to the mass of the universe. Moreover, there may be mass in the universe 
that is not associated with any collapsed structures at all. For 
current generation of simulations
only about 40\% of the total mass has been resolved into halos above 
$10^{11}h^{-1}M_{\sun}$ \cite{2001MNRAS.321..372J}.
It is not obvious that as the resolution of simulations increases 
this fraction converges 
to unity, which is what is commonly assumed in the forms 
of mass function (\citeNP{1974ApJ...187..425P}, \citeNP{1999MNRAS.308..119S}),
since some fraction of the mass could remain in a diffuse form. 
All this makes any extrapolation of $M_{\rm vir}/L$
to the total luminosity density as a method to deduce the
density of the universe highly uncertain. 
One can attempt to correct for this 
using simulations (e.g. \citeNP{2000ApJ...541....1B}), but this relies 
on the ability of these to reproduce the light distribution in the 
universe across a large dynamic range of masses and scales. 
This is one venue to pursue in the future
as simulations and modelling of physical processes improve.

The alternative way is 
to measure mass and light 
over a much larger volume, so that it becomes
representative for the whole universe. Currently there are no reliable
methods that can measure the mean mass directly on such large scales. 
The closest method to achieve this goal is 
gravitational lensing, which measures ellipticity distortions of 
background galaxies. This method however cannot measure the mean component
of the matter, which does not produce shear. 
So instead one must look at fluctuations around the mean by comparing
the relation between galaxy light and convergence 
\shortcite{2001ApJ...556..601W}. 
However, if galaxies are a biased tracer of dark matter then only
a combination $\Omega_m/b$ can be determined and if $b>1$ this will
underestimate $\Omega_m$. 
Since this is a luminosity weighted statistic it will give a larger 
weight to brighter galaxies, which are known to be biased relative 
to fainter ones (e.g. \shortciteNP{2001MNRAS.328...64N}; 
\shortciteNP{2001astro.ph..6476S}).

Since it is difficult to detect a signal on large scales 
with gravitational lensing, where linear biasing applies, one 
must compare the fluctuations in light and mass 
at smaller scales. At any given 
scale halos of a certain mass dominate the fluctuations 
(see e.g. \citeNP{2000MNRAS.318..203S} for a discussion of this in the 
context of halo models). 
For example, on a megaparsec scale the dominant fluctuations come from 
groups and clusters, while on 50kpc scale the galactic halos dominate 
the fluctuations. This 
means that even if the galaxies are unbiased the relation between light 
and mass will be appropriate only for the halos that dominate the fluctuations
at that smoothing scale. As discussed above 
this mass to light relation may not be the 
universal one, since most of the mass may be in smaller objects or 
in diffuse structures, which have too weak signal to be detected through the 
weak lensing fluctuations.
There is no preferred scale at which one should evaluate
$M/L$ to 
multiply it with the total galaxy light to derive the mean 
density of the universe.
If $M/L$ increases 
with halo mass above $L_*$, as suggested by observations 
(\shortciteNP{2001astro.ph.12534G}, 
\citeNP{2002astro.ph..1448G}), then the 
convergence-light correlation function will be more extended than that of 
light itself and there is some observational evidence
for this effect \shortcite{2001ApJ...556..601W}. It is not clear however 
whether this leads to an underestimate or overestimate of $\Omega_m$, 
since the trend of $M/L$ to increase with mass is likely to be 
reversed below $L_*$ and a lot mass associated with 
either diffuse structures or small halos may not be associated with any 
light at all.

Above arguments suggest that global mass to light ratio 
method of determining the density of the 
universe cannot be derived without making additional assumptions.
In the approach presented here we instead
limit the analysis only to the galaxies around $L_*$ which 
are well studied.
We combine the virial mass to luminosity relations extracted 
from the SDSS data with the SDSS luminosity function of $L_*$ galaxies 
and compare that to the universal mass function
to place limits on cosmological models. 
The analysis is done entirely within 
the SDSS data set, which reduces the uncertainties related to the
photometric calibrations and color transformations, which usually 
plague luminosity function comparisons.

\section{Relation between galaxy and halo abundances}
The halo mass function describes the number density of halos
as a function of mass. It can be written as
\begin{equation}
{dn \over d\ln M} ={\bar{\rho} \over M}f(\sigma){d \ln \sigma^{-1} \over 
d \ln M},
\end{equation}
where $\bar{\rho}$ is the mean matter density of the universe, $M$ is the 
virial mass of the halo and $n(M)$ is the spatial number density of halos 
of a given mass $M$. We
introduced a
function $f(\sigma)$, which 
has a universal form independent of the
power spectrum, matter density, normalization or redshift if written
as a function of rms variance of linear density field, 
\begin{equation}
\sigma^2(M)=4\pi \int P(k)W_R(k)k^2 dk.
\end{equation}
Here $W_R(k)$ is the Fourier transform
of the spherical top hat window with radius $R$, chosen such that it encloses 
the mass $M=4\pi R^3 \bar{\rho}/3$ and $P(k)$ is the linear power spectrum.

\begin{figure}
\begin{center}
\leavevmode
\epsfxsize=3.0in \epsfbox{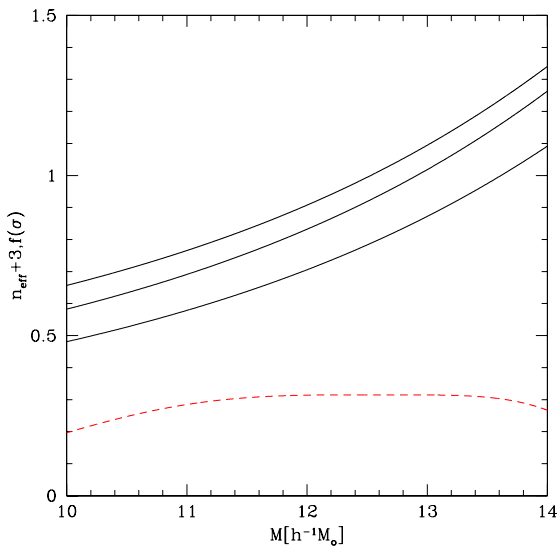}
\end{center}
\caption{$n_{\rm eff}+3$ for models with $\Omega_m=0.3$, $\Omega_b=0.04$, 
$h=0.7$, $n=1$ (solid, 
top), $n=0.9$ (solid, middle) and $\Omega_m=0.25$, $\Omega_b=0.04$, 
$h=0.65$,  
$n=0.8$ (solid, bottom). 
All the transfer functions were computed using CMBFAST.
Also shown is $f[\sigma(M)]$ (dashed), which is essentially 0.3
over this range of masses.
}
\label{fig1}
\end{figure}

The relation between mass and rms variance depends on the linear power 
spectrum. For a smooth power spectrum one can locally approximate 
it as $P(k) \propto k^{n_{\rm eff}}$ and the relation 
is 
\begin{equation}
{d \ln \sigma^{-1} \over d \ln M}={n_{\rm eff}+3 \over 6} .
\end{equation}
For CDM models on galactic scales the effective slope  
$n_{\rm eff} $ ranges between -1.5 and -2.5 (figure \ref{fig1}). 

The universality of the
mass function has been recently investigated by a number 
of authors (\citeNP{1999MNRAS.308..119S}, \citeNP{2001MNRAS.321..372J}, 
\citeNP{2001A&A...367...27W}). It has been shown that
the mass function is indeed universal for a broad range of cosmological 
models. \shortciteN{2001MNRAS.321..372J}
propose the following form,
\begin{equation}
f(\sigma)=0.315 \exp[-|\ln \sigma^{-1}+0.61|^{3.8}],
\end{equation}
which they argue is universal if mass is expressed in terms of the 
virial radius where overdensity is 200 in units of mean density.
It is remarkable that the mass function in this form is almost constant 
for all halos with $\ln \sigma>0$. 
Here we are interested in halos on galactic scales where $\ln \sigma \sim 1$,
so this limit applies and we can take 
$f\sim 0.3$, a universal value which varies only weakly with 
halo mass, as shown in figure \ref{fig1} (dashed line).

To relate the theoretical halo abundance with the observed galaxy 
abundance we must assume a relation between the galaxy luminosity
and its halo mass. Here we use a
direct probe of the halo virial mass obtained from the 
galaxy-galaxy lensing of SDSS \cite{2001astro.ph..8013M}.
Detailed modelling of CDM profiles 
shows that these observations are best 
fitted with $L_*$ galaxies (which dominate the luminosity 
distribution of that sample) having a virial 
mass around $10^{12}h^{-1}M_{\sun}$ \cite{2002astro.ph..1448G}.
The virial mass strongly depends on the morphology or color: 
for a given luminosity 
early type galaxies can be significantly more massive than late type 
galaxies. The differences are reduced in red bands ($r'$, $i'$ and $z'$),
where we focus our analysis as well, but they can still vary by a 
factor of 2-3. This is to a large extent a consequence of stellar age, 
which modifies the stellar mass to light ratio by a similar factor, with 
early type galaxies having a higher stellar mass to light ratio than 
late type galaxies. Current data are consistent with stellar mass to 
virial mass being independent of stellar age (or morphology).
Nevertheless, in the absence of stellar mass 
information for individual galaxies, one must extract all the relations
separately for  the two types before
adding them together. For example, 
for an $L_*=2.1\times 10^{10}h^{-2}L_{\sun}$ in i' \cite{2001AJ....121.2358B}
one finds virial $M_{200\Omega_m}=4.3 \times 10^{11}h^{-1}M_{\sun}$ for late type 
galaxies and $M_{200\Omega_m}=1.2 \times 10^{12}h^{-1}M_{\sun}$ for early 
type galaxies, where we have corrected for a 25\% increase from $M_{200}$
to $M_{200\Omega_m}$
assuming $\Omega_m=0.3$ fiducial model.

An alternative and more model dependent estimate of virial masses comes 
from the
Tully-Fisher (TF) relation for late type galaxies 
\shortcite{1997ApJ...477L...1G}
or Faber-Jackson (FJ) relation 
for early type galaxies \shortcite{2001astro.ph.10344B}.  
These observe galaxy properties in the inner 10$h^{-1}$kpc and do 
not directly measure the virial mass of the halo in which the galaxy sits. 
At these radii the dynamical effect of baryons on the rotation curves 
(as well as on the dark matter distribution itself) is important and modifies 
the relation between the rotation velocity and virial mass. 
One can however model this assuming stellar mass to light ratio and dark 
matter profile. The first comes from the stellar population synthesis 
models and is rather uncertain because of age and IMF 
\cite{1993ApJ...405..538B}, while
the latter is obtained 
from the cosmological dark matter simulations  and depends on 
the assumed cosmological model \shortcite{2001MNRAS.321..559B}. 
In addition, the response of dark matter 
to baryon cooling must be included and is often 
modelled with adiabatic contraction \shortcite{1986ApJ...301...27B}.
Adopting standard values for the dark matter profiles and the stellar mass
to light ratios one finds that at $L_*$ the rotation velocity 
decreases by 1.8 from optical to virial radius both for early and late 
type galaxies, in good agreement with g-g lensing results \cite{2002astro.ph..1450S}.

Galaxy-galaxy lensing can also be used to determine the slope $\beta$
of the relation between mass and luminosity,
\begin{equation}
{M \over M_*}=\left({L \over L_*}\right)^{\beta},
\end{equation}
where $M_*$ is the mass associated with $L_*$ galaxy. 
Using SDSS data one 
finds $\beta = 1.4 \pm 0.2$ in red bands, which
becomes $\beta^{\rm e} = 1.2 \pm 0.2$ after correcting for a luminosity 
dependent fraction of early type galaxies in the sample 
\cite{2002astro.ph..1448G}. This 
is valid only for 
early type galaxies between $L_*$ and $7L_*$, 
which dominate the g-g lensing signal. 
Note that this differs significantly from FJ relation, which would 
predict $\beta \sim 2/3$, but agrees well with the detailed modelling 
of rotation curves for early type galaxies \cite{2002astro.ph..1450S}. 
To determine $\beta$ around $L_*$ 
for late type galaxies we cannot use g-g lensing, since the signal is 
too weak to be detected as a function of luminosity. Instead we
use TF relation which gives $L_I \propto v_{\rm opt}^{3.1}$ \cite{1997ApJ...477L...1G}.
We model the contributions from stellar disk and 
adiabatic response of dark matter to disk formation to relate between 
the rotation velocity at the optical radius and virial velocity (or mass). 
We assume NFW profile with $c_{200}=10$ and use $\Upsilon_I=1.5h$, 
which were shown to reproduce well the virial velocity constraint from 
g-g lensing at $L_*$ \cite{2002astro.ph..1450S}. 
We find $\beta^{\rm l} \sim  1$ around $L_*$, 
which is the value we adopt below.

The third parameter that we need is the fraction $\gamma$ of galaxies that are at the centers of
isolated galactic halos, as opposed to larger halos
such as groups and clusters. 
Universal mass functions from N-body simulations only count
isolated halos, so one must correct for the fraction of galaxies at
a given luminosity that are in larger groups and clusters.
This fraction can be determined from the relative 
contribution to galaxy-galaxy lensing at small and large separations.
Above 200-300$h^{-1}$kpc the signal is dominated by groups and clusters,
while below it is 
dominated by individual galactic halos. Analysis of SDSS galaxy-galaxy 
lensing data finds $\gamma=0.72\pm 0.1$ for early type galaxies and
$\gamma=0.93 \pm 0.1$ for late type galaxies around $L_*$ 
\cite{2002astro.ph..1448G}. 
The error includes various systematic uncertanties, of which 
the galaxy occupation as a function of group and cluster mass is the most 
important. Thus about 10-30\% 
of $L_*$ galaxies reside in groups and clusters as defined in 
N-body simulations and one must reduce the 
galaxy abundance by this fraction when relating it to the halo abundance. 

The abundance of galaxies of a given luminosity can be extracted from 
the luminosity function $dn(L)/d\ln L$, 
which determines the abundance per logarithmic interval of
 luminosity. It is often fitted to the Schechter form,
\begin{equation}
\Phi(L)\equiv {dn \over d \ln L}=\phi_*\left({L \over L_*}\right)^{\alpha+1}\exp(-L/L_*).
\label{lf}
\end{equation}
For $L_*$ galaxies the abundance is $\Phi(L_*)=\phi_*/e$, where 
$e=2.718$ is the natural logarithmic base constant. From the early SDSS 
analysis the values for $\phi_*$
are $1.46\times 10^{-2}$, $1.28\times 10^{-2}$ and $1.27\times 10^{-2}$
(with a 10\% error)
in units of $h^{3}{\rm Mpc}^{-3}$ for $r'$, $i'$ and $z'$, respectively, 
while the values for $\alpha$ are around -1.2 to -1.25 with a few percent 
error \cite{2001AJ....121.2358B}. We 
choose these 3 bands because they show least variation in virial 
mass to light 
ratio between early and late type galaxies, minimizing the 
color dependence of the signal. In addition, the redshift
evolution corrections in red bands are smaller than in $g'$ or $u'$. 
These have been suggested 
as one possible reason for the discrepancy between the 2dF luminosity 
function in $b_J$ and SDSS in $g'$ \shortcite{2001astro.ph.11011N}. 

The last ingredient that is needed is the fraction of  
early type ($\xi^{\rm e}$) and late type ($\xi^{\rm l}$) 
galaxies as a function of magnitude. As shown in 
\citeN{2001astro.ph..7201S}  
this fraction depends on luminosity, so that the early type galaxies
dominate at the bright end and the late type galaxies dominate at the faint end.
This statement is only valid for red bands and a reverse trend is seen 
in $u'$, while $g'$ shows comparable fractions almost independent of 
luminosity. 
In red bands 
around $L_*$ the fraction is somewhat higher for the
early types.
While the transition between the early and late types is not well 
defined the transitional types (S0-Sa) do not dominate the counts, so here
we will simply assume a bimodal early/late 
distribution ($\xi^{\rm e}+\xi^{\rm l}$=1),
rather than attempt to model the whole range of stellar ages and 
morphologies. 
This could be improved in the future as larger statistical samples 
are obtained and is particularly important for spirals, 
which have a larger scatter in the stellar ages.
Early type galaxies are more homogeneous and old, so their scatter 
in stellar mass to light ratio should be small.

We can now combine the above equations to relate halo and galaxy 
abundances. This gives
\begin{equation}
{dn \over d\ln M}={dn \over \beta d\ln L}= {\gamma\Phi(L) \over \beta}=
{n_{\rm eff}+3 \over 6}{\bar{\rho} \over M_{200 \Omega_m}}f.
\end{equation}
Using $\bar{\rho}=\Omega_m \rho_{\rm crit}$=$2.77\times 10^{11}h^2M_{\sun}
{\rm Mpc}^{-3}$ 
one finds the minimum mean density at a given luminosity is given by
\begin{equation}
\Omega_m(n_{\rm eff}+3)= 0.3 {\xi \gamma \over f\beta}
\left({ \Phi(L)M_{200 \Omega_m} \over 
1.4\times 10^{10}h^{2}M_{\sun}{\rm Mpc}^{-3}}\right).
\label{omegam}
\end{equation}
Note that one must add up the contribution from both early and late type 
galaxies separately, 
where the two contributions have to be evaluated at equal mass, not 
luminosity. If there are dark halos without a bright galaxy at the 
center the above expression become inequality and one can only 
place a lower limit on $\Omega_m(n_{\rm eff}+3)$.
We parametrize this uncertainty with the fraction of dark 
halos $f_{\rm dh}$, which in general is a function of halo mass.

We can evaluate this expression at several different values for 
halo mass. The virial mass of an early type 
galaxy at $L_*=2.1\times 10^{10}h^{-2}L_{\sun}$ in $i'$
is $M_{200 \Omega_m}=1.2\times 10^{12}h^{-1}M_{\sun}$ \cite{2002astro.ph..1448G}. Using
$\beta^{\rm e}= 1.2$, $\gamma^{\rm e}=0.72$ and $\xi^{\rm e}=0.6$ one finds
$\Omega_{\rm m}^{\rm e}(n_{\rm eff}+3)=0.14/(1-f_{\rm dh})$.
To this we must add the 
contribution from late types at the same mass. At $L_*$ their virial mass is 
$M_{200 \Omega_m}=4.3\times 10^{11}h^{-1}M_{\sun}$. Using
$\beta^{\rm l} = 1$ one finds that for 
$M_{200 \Omega_m}=1.2\times 10^{12}h^{-1}M_{\sun}$ 
the corresponding 
luminosity is $3L_*$. At this luminosity the fraction of late type galaxies
in the sample is already small, $\xi^{\rm l}\sim 0.2$. In addition, 
from the luminosity function in equation (\ref{lf})
one can see the abundance of $3L_*$ galaxies
has decreased by a factor of 10 relative to that of $L_*$. So late 
type galaxies do not actually add much to the limit above and together 
we find $\Omega_{\rm m}(n_{\rm eff}+3)=0.15/(1-f_{\rm dh})$. 
These constraints are evaluated in $i'$, but one finds similar 
constraints also in $r'$ and $z'$. The error budget is dominated by 
the errors on $M_{200\Omega_m}$, $\gamma$
and $\beta$, which combined give about 30\% 
uncertainty.

We can repeat the same analysis one magnitude above and below $L_*$. 
At $L=2.5L_*=5.2\times10^{10}h^{-2}L_{\sun} $ 
the sample is dominated by early type galaxies, 
so $\xi^{\rm e}\sim 0.8$, $\beta^{\rm e} \sim 1.2$ and $M=3.6 
\times 10^{12}h^{-1}M_{\sun}$. The fraction of these 
galaxies in isolated galactic halos 
is not well determined, but is likely to be larger than 
at $L_*$ \cite{2002astro.ph..1448G}, so we will 
assume $\gamma^{\rm e}=0.9$. This gives $\Omega_{\rm m}
(n_{\rm eff}+3)>0.13/(1-f_{\rm dh})$, 
where we have ignored the very small contribution from the 
late type galaxies. The effective slope is about 5-10\%
higher than at $1.2\times 10^{12}h^{-1}M_{\sun}$ (figure \ref{fig1}),
so the obtained value is about 20\% lower 
than the value obtained at $L_*$. If we assumed 
$\gamma^{\rm e}$ does not differ from that at $L_*$ we find the two 
estimates are in perfect agreement. This 
is quite impressive given that the masses and abundances 
change by a factor of several. The error is comparable to the error 
at $L_*$.

One magnitude below $L_*$ the sample is dominated by late type galaxies, 
for which we use 
$\beta^{\rm l} \sim 1$, $M=1.7\times 10^{11}h^{-1}M_{\sun}$ and 
$\gamma^{\rm l} \sim 0.9$. Adopting $\xi^{\rm l}=0.8$ leads to  
$\Omega_m(n_{\rm eff}+3)=0.11$. To this we have to add the contribution from 
early types at $M=1.7\times 10^{11}h^{-1}M_{\sun}$. 
If $\beta^{\rm e}=1.2$ extends to this mass range this mass 
corresponds to $L \sim 0.1L_*$ and at this luminosity the fraction of 
early type galaxies is about 10\%. This
increases the above estimate by 20\%, so 
$\Omega_m(n_{\rm eff}+3)=0.13/(1-f_{\rm dh})$.
This estimate
is more uncertain, since both
$\beta^{\rm l,e}$ and $\gamma^{\rm l,e}$ have not been directly measured 
over this range.
In addition, the large scatter in stellar ages for late type spirals 
leads to a scatter in mass to luminosity relation.
Note that at this mass one expects $n_{\rm eff}+3$ to decrease by about 20\%
relative to $1.2\times 10^{12}h^{-1}M_{\sun}$ (figure \ref{fig1}),
so this constraint is actually very similar to the one at $L_*$ based 
on early type galaxies, even though the masses differ by almost an 
order of magnitude.

From the above analysis we find that over the 
range of masses between $1.6\times 10^{11}h^{-1}M_{\sun}$ to 
$3.6\times 10^{12}h^{-1}M_{\sun}$ the cosmological constraints 
on $\Omega_m/(1-f_{\rm dh})$ are all very similar, 
\begin{equation}
\Omega_{\rm m}(n_{\rm eff}+3)(1-f_{\rm dh})=0.15\pm 0.05 .
\end{equation}
The good agreement found over a wide range of mass suggests that 
the shape of the 
mass function agrees well with the one predicted from cosmological 
simulations, assuming the fraction of dark halos $f_{\rm dh}$, 
if different from 
zero, does not vary over this mass range.

\section{Discussion}
In this paper we propose a method to derive cosmological constraints from
the virial masses and abundance of $L_*$ galaxies and apply it to 
early SDSS observations. The method 
differs from other methods using mass to light ratio $M/L$ 
in that it only uses this
information around $L_*$ galaxies and not the overall luminosity density. 
This sidesteps the uncertainties related to the variation of $M/L$ with 
luminosity $L$. 
The main observational inputs are relation between virial mass and luminosity
around $L_*$ as 
a function of morphological type and the fraction of these galaxies in 
isolated halos as opposed to groups and clusters, 
all of which can be extracted from g-g
lensing. Another essential ingredient is
the luminosity function of galaxies around $L_*$ as a function of 
morphological type, which can be obtained from the same data as g-g 
lensing information.
The main theoretical input is the halo mass function, which has been 
shown to be universal by a number of studies 
(\citeNP{1999MNRAS.308..119S}, \citeNP{2001MNRAS.321..372J},
\citeNP{2001A&A...367...27W}).
The abundance of halos depends only on the density 
parameter $\Omega_m$ and the effective slope of the linear power 
spectrum $n_{\rm eff}$ through the combination $\Omega_m(n_{\rm eff}+3)$. 

In the absence of dark halos 
the obtained constraint $\Omega_m(n_{\rm eff}+3)=0.15 \pm 0.5$ is 
low compared to the 
predictions of $\Lambda$CDM model with $\Omega_m=0.3$, $h=0.7$ 
and $n=1$, which gives $\Omega_m (n_{\rm eff}+3)=0.28$ at these scales. 
This is excluded by the current constraints,
unless a significant fraction of 
halos is dark over this range of masses. 
To bring the models into a better agreement with this constraint 
one can either lower 
the effective slope or the mean density.
The former can be lowered by reducing the primordial spectrum slope $n$, 
decreasing the shape parameter $\Gamma$ 
(which depends on $\Omega_m=\Omega_{dm}+\Omega_b$, $\Omega_b$ and
Hubble constant $H_0$, see e.g.
\citeNP{1998ApJ...496..605E}) or introducing warm 
dark matter \cite{2001ApJ...556...93B}. For example, we find that a model 
with $\Omega_m=0.25$, $h=0.65$, $\Omega_b=0.04$ and $n=0.8$ 
gives a good agreement with the observational constraints, but somewhat 
lower values of $\Omega_m \sim 0.2$ with $n=0.9-1.0$ are also acceptable.
For warm dark matter models, which suppress power on small scales,
we compute transfer functions
using CMBFAST \cite{1996ApJ...469..437S} and find that the effective slope at the
galactic mass scale remains almost unchanged for $m_{\nu}>500eV$ and drops to 
$n_{\rm eff}=-2.5$ at $m_{\nu}=250eV$. 
Such low masses are probably excluded from $Ly-\alpha$ forest studies
\cite{2000ApJ...543L.103N}, although this must be confirmed with a more careful error analysis
of $Ly-\alpha$ forest constraints.

There are several possible sources of uncertainty that can affect the 
current constraints derived above. 
First there is the possibility that the virial masses used here are too low.
This is certainly
possible for late type galaxies, which have a weak signal in
g-g lensing and for which a factor of 2 increase in mass is less than 
a 2-$\sigma$ excursion. For early type galaxies the statistical error on 
the virial mass is 20\%, so a factor of 2 excursion is unlikely, although 
some systematic uncertainties remain in the g-g lensing analysis.
The 
agreement between early and late type galaxies
implies that the mass scale must be changed 
for both types. Furthermore, any increase in the virial mass
would affect the agreement between g-g lensing masses and 
Tully-Fisher or Faber-Jackson relation. 
A change in mass by a factor of 2
would reduce the optical rotation velocity 
to virial rotation velocity ratio from 1.8 to 1.4 \cite{2002astro.ph..1450S}.
Within the context of adiabatic compression models such a ratio 
is likely to be too small to be explained with CDM profiles and stellar
mass to light ratios as expected from stellar population 
synthesis models with reasonable IMF. This possibility would thus require
one to give up a successful prediction of CDM models, that of the structure 
of dark matter halos in the outer parts.

Another possibility is that the
relation between the halo mass and luminosity $M\propto L^{\beta}$
is shallower than assumed here, $\beta \sim 0.6$. 
This would 
be the case if virial mass to light ratio was decreasing with luminosity.
Both g-g lensing and modelling of optical relations are consistent
with $\beta=1-1.5$ at $L_*$ and above and theoretical models also 
predict $\beta>1$ in this regime (e.g. \shortciteNP{1999MNRAS.303..188K}, \shortciteNP{2000MNRAS.311..793B}). While $M/L$ probably decreases with $L$
well below $L_*$, this is unlikely to be the case over the range of luminosities
of interest here. If $\beta^{\rm e}=1$ instead of 1.2 used here this 
would increase the lower limit on $\Omega_m(n_{\rm eff}+3)$ by 20\%. 
It seems similarly unlikely that the luminosity 
function can be wrong by more than 30\%, unless there is a large 
fraction of low surface brightness galaxies that are missed by the SDSS
survey. There are still important calibration differences between 
different surveys, which can cause a mismatch in derived luminosity functions 
\shortcite{2001astro.ph.11011N}, 
but these are not relevant for our analysis since we derive the g-g lensing
relation between light and luminosity using the same sample  
that is used for luminosity function as well.

Another uncertainty is the division into early and late type galaxies and 
their associated fractions, which is somewhat artificial, since there is a
continuous range of colors and light concentrations 
in the sample. What is really 
relevant is the stellar mass of the galaxy, which seems to correlate
quite well with the virial mass. For late type galaxies the range of 
stellar ages is large and this introduces a spread in the luminosity 
for a given stellar mass, 
so representing them with a single mass to light ratio may not be accurate.
This is less of an issue for early type galaxies, which are very 
old and for which the spread in the stellar mass to light ratio is small. 
On the other hand, early type galaxies tend to reside in 
denser enviroments and the fraction of these in groups and clusters is
larger. This correction is also somewhat uncertain, since one must assume
how the groups and clusters are populated with these galaxies to 
determine it \cite{2002astro.ph..1448G}. 
There is also the issue whether the mass
profile of galaxies within groups and clusters
differs from that of the same luminosity 
in the field within inner 100kpc. Observations \cite{2002astro.ph..1448G}
and numerical simulations \shortcite{2000ApJ...544..616G} suggest they do 
not differ much, but the uncertainties are still large. 
The extreme case is when there is 
actually no mass attached to individual galaxies inside groups and 
clusters. Then one should 
use $\gamma^{\rm e}=1$, which would increase the estimated 
$\Omega_m(n_{\rm eff}+3)$ by 30\%.

The choice of the virial radius and 
its associated mass, defined differently by different authors,
has a minor effect on our results. 
From the simulations it is usually defined by friends-of-friends
algorithm (assuming a specific value of linking parameter, e.g. 0.2),
but it is not always clear how this relates to the specific mean density within
the virial radius. For consistency with \citeN{2001MNRAS.321..372J}
we define the virial radius as the value where the mean overdensity is 
$\bar{\delta}=200$, while the observed masses are usually expressed 
in terms of overdensity relative to the critical density. 
For a given halo profile one can convert between the two, but the 
conversion depends on the assumed density parameter $\Omega_m$. 
Fortunately the differences
are not very important on galactic scales, where halos 
are highly concentrated and 
where the mass around the virial radius only slowly grows with radius 
(logarithmically in the limit of a large concentration where the slope at 
virial radius approaches -3). 
For example, the difference in mass
between $\bar{\delta}=200\Omega_m$ for $\Omega_m=0.4$ and $\Omega_m=0.2$ 
in a halo of
$c=12$ is 10\%, so this is not the dominating source of error.
This effect has been included in the estimates above.

The remaining uncertainty is the fraction of dark halos in the 
universe. Theoretically one would not expect halos to be dark 
around $L_*$, where the efficiency of cooling and star formation is 
high. 
This is supported by the fact that for the halos that we 
do see a large fraction of the baryons within the virial radius has converted 
into stars \cite{2002astro.ph..1448G}. 
At lower halo masses, below $10^{11}h^{-1}M_{\sun}$,
 the fraction of dark halos may increase 
because some of the formed disks may be stable against star formation 
\cite{2002astro.ph..2283V}. At higher halo masses above $10^{13}h^{-1}M_{\sun}$ 
one enters the group regime, cooling becomes less efficient,
and a significant
fraction of these halos may not host a bright galaxy at the center.
For example, applying the same analysis as in this paper to
7$L_*$ early type galaxies one finds the 
abundance of halos to be several times below that 
expected from the mass function, 
a consequence of exponentially decreasing luminosity function at the 
bright end. This is not surprising, since a
large fraction of the groups in this range probably hosts
several fainter galaxies rather than a single bright galaxy at the center.
It is difficult to determine the abundance of such groups directly 
from the optical data, since it is not obvious which groups are virialized
to satisfy the halo definition in an N-body simulation and which are just 
a collection of galaxies approaching each other for the first time (as for
example the local group). The fact that the constraints derived here 
agree from $1.6\times 10^{11}h^{-1}M_{\sun}$ to 
$3.6\times 10^{12}h^{-1}M_{\sun}$ suggests that 
over this range of masses the simplest possibility with $f_{\rm dh}=0$
is consistent with the data.

The prospects to determine the fraction of dark halos directly
seem difficult. The only direct way to observe these is 
through the gravitational lensing, but in the absence of a significant 
baryon condensation such halos are inefficient strong lenses \cite{2001ApJ...559..531K}. 
The halos that do cool and form a disk without making stars
could be somewhat more efficient for a given halo mass, 
but are expected to have lower halo masses. Most of the lenses
are bright early type galaxies which both reside in massive halos and 
have a significant baryonic contribution to the lensing cross-section.
It is thus possible that even if some fraction of halos around
$10^{12}h^{-1}M_{\sun}$ exists without a central galaxy 
they may not have been detected so far with strong lensing. 
Current surveys such as SDSS may provide better limits on the fraction 
of dark halos as a function of halo mass. 

It is clear from the above discussion that the errors in the current
analysis are still quite large, although the fact that the constraints 
are consistent over a range of masses increases the confidence level 
of the final result. 
It is interesting that the constraints obtained 
are in a good agreement with the cluster gas fraction determination of 
the matter density \cite{2002astro.ph..2357E} and with the redshift
distortions and bias determination from 2dF \cite{2001astro.ph.12161V}. 
They are also 
comparable or somewhat higher than those using global $M/L$ ratio
(\shortciteNP{2000ApJ...541....1B}, \shortciteNP{2001ApJ...556..601W}), 
although this method, as discussed above, may well be biased. 
Similarly, a tilted CDM model may help aleviate some of the small 
scale problems with CDM \cite{2001astro.ph..9392K}. 
This perhaps indicates the need for a somewhat 
lower $\Omega_m$ or $n_{\rm eff}$ 
than previously suggested $\Omega_m=0.3-0.4$, $n=1$ model. 

While the error from the method presented here
is still large, the prospects to improve it are
good. Currently the errors are dominated by observationally 
determined parameters $M_{200\Omega_m}$, $\phi_*$, $\beta$ and $\gamma$. 
These errors are dominated by statistics and 
were obtained by using only 5\% of the final SDSS sample. 
With the full sample one can reduce the error on each of these 
significantly, as well as extend the observable range to the lower luminosity 
galaxies. With the full sample one can also study 
the morphological dependence of the mass-luminosity relation in more
detail, which as we have shown here plays an important role in the analysis.
With spectroscopic information it should be able to 
extract stellar mass information for each galaxy separately and 
use g-g lensing to relate the stellar mass to the virial mass directly 
without splitting the sample into morphological types as done here.
Theoretical errors are mainly caused by the 
accuracy of the mass function over this mass range, but  
the uncertainty is already at a 10\% level and
can be further improved with simulations that cover a wider range 
of cosmological models. The remaining uncertainty is the fraction of 
dark halos, which should also be 
determined with a comparable accuracy (or shown to be negligible). In this
case the method presented here may
become an accurate test of matter density and slope of the power spectrum 
on one megaparsec scale. 

The author acknowledges the support of NASA, David and Lucille
Packard Foundation and Alfred P. Sloan Foundation.
I thank Neta Bahcall, David Spergel and Simon White for useful comments.

     \bibliography{apjmnemonic,cosmo,cosmo_preprints}

\begin{thebibliography}{}

\bibitem[\protect\citeauthoryear{{Alam}, {Bullock}, \& {Weinberg}}{{Alam}
  et~al.}{2001}]{2001astro.ph..9392K}
{Alam} K.~S.~M., {Bullock} J.~S.,  {Weinberg} D.~H., 2001, in astro-ph/0109392

\bibitem[\protect\citeauthoryear{{Bahcall} et~al.}{{Bahcall}
  et~al.}{2000}]{2000ApJ...541....1B}
{Bahcall} N.~A., {Cen} R., {Dav{\' e}} R., {Ostriker} J.~P.,  {Yu} Q., 2000,
  \apj, 541, 1

\bibitem[\protect\citeauthoryear{{Bahcall}, {Lubin}, \& {Dorman}}{{Bahcall}
  et~al.}{1995}]{1995ApJ...447L..81B}
{Bahcall} N.~A., {Lubin} L.~M.,  {Dorman} V., 1995, \apjl, 447, L81

\bibitem[\protect\citeauthoryear{{Benson} et~al.}{{Benson}
  et~al.}{2000}]{2000MNRAS.311..793B}
{Benson} A.~J., {Cole} S., {Frenk} C.~S., {Baugh} C.~M.,  {Lacey} C.~G., 2000,
  \mnras, 311, 793

\bibitem[\protect\citeauthoryear{{Bernardi} et~al.}{{Bernardi}
  et~al.}{2001}]{2001astro.ph.10344B}
{Bernardi} M. et~al., 2001, in astro-ph/0110344

\bibitem[\protect\citeauthoryear{{Blanton} et~al.}{{Blanton}
  et~al.}{2001}]{2001AJ....121.2358B}
{Blanton} M.~R. et~al., 2001, \aj, 121, 2358

\bibitem[\protect\citeauthoryear{{Blumenthal} et~al.}{{Blumenthal}
  et~al.}{1986}]{1986ApJ...301...27B}
{Blumenthal} G.~R., {Faber} S.~M., {Flores} R.,  {Primack} J.~R., 1986, \apj,
  301, 27

\bibitem[\protect\citeauthoryear{{Bode}, {Ostriker}, \& {Turok}}{{Bode}
  et~al.}{2001}]{2001ApJ...556...93B}
{Bode} P., {Ostriker} J.~P.,  {Turok} N., 2001, \apj, 556, 93

\bibitem[\protect\citeauthoryear{{Bruzual} \& {Charlot}}{{Bruzual} \&
  {Charlot}}{1993}]{1993ApJ...405..538B}
{Bruzual} G.,  {Charlot} S., 1993, \apj, 405, 538

\bibitem[\protect\citeauthoryear{{Bullock} et~al.}{{Bullock}
  et~al.}{2001}]{2001MNRAS.321..559B}
{Bullock} J.~S., {Kolatt} T.~S., {Sigad} Y., {Somerville} R.~S., {Kravtsov}
  A.~V., {Klypin} A.~A., {Primack} J.~R.,  {Dekel} A., 2001, \mnras, 321, 559

\bibitem[\protect\citeauthoryear{{Carlberg}, {Yee}, \& {Ellingson}}{{Carlberg}
  et~al.}{1997}]{1997ApJ...478..462C}
{Carlberg} R.~G., {Yee} H.~K.~C.,  {Ellingson} E., 1997, \apj, 478, 462

\bibitem[\protect\citeauthoryear{{Eisenstein} \& {Hu}}{{Eisenstein} \&
  {Hu}}{1998}]{1998ApJ...496..605E}
{Eisenstein} D.~J.,  {Hu} W., 1998, \apj, 496, 605

\bibitem[\protect\citeauthoryear{{Erdogdu}, {Ettori}, \& {Lahav}}{{Erdogdu}
  et~al.}{2002}]{2002astro.ph..2357E}
{Erdogdu} P., {Ettori} S.,  {Lahav} O., 2002, in 7 pages, 5 figures, submitted
  to MNRAS.

\bibitem[\protect\citeauthoryear{{Ghigna} et~al.}{{Ghigna}
  et~al.}{2000}]{2000ApJ...544..616G}
{Ghigna} S., {Moore} B., {Governato} F., {Lake} G., {Quinn} T.,  {Stadel} J.,
  2000, \apj, 544, 616

\bibitem[\protect\citeauthoryear{{Giovanelli} et~al.}{{Giovanelli}
  et~al.}{1997}]{1997ApJ...477L...1G}
{Giovanelli} R., {Haynes} M.~P., {da Costa} L.~N., {Freudling} W., {Salzer}
  J.~J.,  {Wegner} G., 1997, \apjl, 477, L1

\bibitem[\protect\citeauthoryear{{Girardi} et~al.}{{Girardi}
  et~al.}{2001}]{2001astro.ph.12534G}
{Girardi} M., {Manzato} P., {Mezzetti} M., {Giuricin} G.,  {Limboz} F., 2001,
  in astro-ph/0112534

\bibitem[\protect\citeauthoryear{{Gonzalez} et~al.}{{Gonzalez}
  et~al.}{2000}]{2000ApJ...528..145G}
{Gonzalez} A.~H., {Williams} K.~A., {Bullock} J.~S., {Kolatt} T.~S.,  {Primack}
  J.~R., 2000, \apj, 528, 145

\bibitem[\protect\citeauthoryear{{Guzik} \& {Seljak}}{{Guzik} \&
  {Seljak}}{2002}]{2002astro.ph..1448G}
{Guzik} J.,  {Seljak} U., 2002, in astro-ph/0201448

\bibitem[\protect\citeauthoryear{{Jenkins} et~al.}{{Jenkins}
  et~al.}{2001}]{2001MNRAS.321..372J}
{Jenkins} A., {Frenk} C.~S., {White} S.~D.~M., {Colberg} J.~M., {Cole} S.,
  {Evrard} A.~E., {Couchman} H.~M.~P.,  {Yoshida} N., 2001, \mnras, 321, 372

\bibitem[\protect\citeauthoryear{{Jimenez} et~al.}{{Jimenez}
  et~al.}{1997}]{1997MNRAS.292L...5J}
{Jimenez} R., {Heavens} A.~F., {Hawkins} M.~R.~S.,  {Padoan} P., 1997, \mnras,
  292, L5

\bibitem[\protect\citeauthoryear{{Kauffmann} et~al.}{{Kauffmann}
  et~al.}{1999}]{1999MNRAS.303..188K}
{Kauffmann} G., {Colberg} J.~M., {Diaferio} A.,  {White} S.~D.~M., 1999,
  \mnras, 303, 188

\bibitem[\protect\citeauthoryear{{Kochanek}}{{Kochanek}}{2001}]{2001astro.ph..%
8160K}
{Kochanek} C.~S., 2001, in the proceedings of The Dark Universe meeting at
  STScI, April 2-5, 2001 M. Livio, ed., Cambridge University Press

\bibitem[\protect\citeauthoryear{{Kochanek} \& {White}}{{Kochanek} \&
  {White}}{2001}]{2001ApJ...559..531K}
{Kochanek} C.~S.,  {White} M., 2001, \apj, 559, 531

\bibitem[\protect\citeauthoryear{{McKay} et~al.}{{McKay}
  et~al.}{2001}]{2001astro.ph..8013M}
{McKay} T.~A. et~al., 2001, in astro-ph/0108013

\bibitem[\protect\citeauthoryear{{Narayanan} et~al.}{{Narayanan}
  et~al.}{2000}]{2000ApJ...543L.103N}
{Narayanan} V.~K., {Spergel} D.~N., {Dav{\' e}} R.,  {Ma} C., 2000, \apjl, 543,
  L103

\bibitem[\protect\citeauthoryear{{Norberg} et~al.}{{Norberg}
  et~al.}{2001}]{2001MNRAS.328...64N}
{Norberg} P. et~al., 2001, \mnras, 328, 64

\bibitem[\protect\citeauthoryear{{Norberg}, {Cole}, \& {the 2dFGRS
  Team}}{{Norberg} et~al.}{2001}]{2001astro.ph.11011N}
{Norberg} P., {Cole} S.,  {the 2dFGRS Team} , 2001, in astro-ph/0111011

\bibitem[\protect\citeauthoryear{{Press} \& {Schechter}}{{Press} \&
  {Schechter}}{1974}]{1974ApJ...187..425P}
{Press} W.~H.,  {Schechter} P., 1974, \apj, 187, 425

\bibitem[\protect\citeauthoryear{{Seljak}}{{Seljak}}{2000}]{2000MNRAS.318..203%
S}
{Seljak} U., 2000, \mnras, 318, 203

\bibitem[\protect\citeauthoryear{{Seljak}}{{Seljak}}{2002}]{2002astro.ph..1450%
S}
{Seljak} U., 2002, in astro-ph/0201450

\bibitem[\protect\citeauthoryear{{Seljak} \& {Zaldarriaga}}{{Seljak} \&
  {Zaldarriaga}}{1996}]{1996ApJ...469..437S}
{Seljak} U.,  {Zaldarriaga} M., 1996, \apj, 469, 437

\bibitem[\protect\citeauthoryear{{Sheth} \& {Tormen}}{{Sheth} \&
  {Tormen}}{1999}]{1999MNRAS.308..119S}
{Sheth} R.~K.,  {Tormen} G., 1999, \mnras, 308, 119

\bibitem[\protect\citeauthoryear{{Strateva} et~al.}{{Strateva}
  et~al.}{2001}]{2001astro.ph..7201S}
{Strateva} I., {Ivezic} Z., {Knapp} G.~R., {Narayanan} V.~K., {Strauss} M.~A.,
  {et al.} , 2001, in astro-ph/0107201

\bibitem[\protect\citeauthoryear{{Verde} et~al.}{{Verde}
  et~al.}{2001}]{2001astro.ph.12161V}
{Verde} L. et~al., 2001, in astro-ph/0112161

\bibitem[\protect\citeauthoryear{{Verde}, {Oh}, \& {Jimenez}}{{Verde}
  et~al.}{2002}]{2002astro.ph..2283V}
{Verde} L., {Oh} S.~P.,  {Jimenez} R., 2002, in astro-ph/0202283

\bibitem[\protect\citeauthoryear{{White}}{{White}}{2001}]{2001A&A...367...27W}
{White} M., 2001, \aap, 367, 27

\bibitem[\protect\citeauthoryear{{Wilson}, {Kaiser}, \& {Luppino}}{{Wilson}
  et~al.}{2001}]{2001ApJ...556..601W}
{Wilson} G., {Kaiser} N.,  {Luppino} G.~A., 2001, \apj, 556, 601

\bibitem[\protect\citeauthoryear{{Zehavi} et~al.}{{Zehavi}
  et~al.}{2001}]{2001astro.ph..6476S}
{Zehavi} I., {Blanton} M.~R., {Frieman} J.~A., {Weinberg} D.~H., {Mo} H.~J.,
  {Strauss} M.~A.,  {et al.} , 2001, in astro-ph/0106476

\end{thebibliography}
	\bibliographystyle{mnras}

\end{document}